# The $K$-Receiver Broadcast Channel with Confidential Messages

Li-Chia Choo and Kai-Kit Wong, *Senior Member, IEEE*

**Abstract**

The secrecy capacity region for the $K$-receiver degraded broadcast channel (BC) is given for confidential messages sent to the receivers and to be kept secret from an external wiretapper. Superposition coding and Wyner's random code partitioning are used to show the achievable rate tuples. Error probability analysis and equivocation calculation are also provided. In the converse proof, a new definition for the auxiliary random variables is used, which is different from either the case of the 2-receiver BC without common message or the $K$-receiver BC with common message, both with an external wiretapper; or the $K$-receiver BC without a wiretapper.

## I. Introduction

The wireless communications channel is vulnerable to eavesdropping or wiretapping due to the open nature of the channel. An important requirement for wireless systems today is the characterization of transmission rates that allow for both secure and reliable communication for the physical layer. Recent studies addressing this issue have included wireless network building blocks such as the multiple-access channels [1], relay channels [2], fading channels [3], [4] and multiple-input multiple-output (MIMO) channels [5].

A suitable model for studying such simultaneously secure and reliable communication in the wireless broadcast and communications medium is the broadcast channel (BC) with confidential messages, which was studied by Csiszár and Körner [6]. It is a generalization of the characterization of the wiretap channel by Wyner [7]. In the model of [6], there are 2 receivers and a common message is sent to both, while a confidential message is sent to one of the receivers and is to be kept secret from the other receiver. The secrecy level is determined by using the equivocation rate, which is the entropy rate of the confidential message conditioned on the channel output at the eavesdropper or wiretapper. The secrecy capacity region is the set of transmission rates where the legitimate receiver decodes its confidential message while keeping the message secret from the wiretapper.

In more recent studies on the BC with confidential messages, Liu *et al.* [8] investigated the scenario where there are 2 receivers and private messages are sent to each one and kept secret from the unintended receiver; Xu *et al.* [9] studied the case in [8] but with a common message to both receivers; Bagherikaram *et al.* [10] addressed the scenario where there are 2 receivers and one wiretapper, with confidential messages sent to the receivers.

The authors are with the Department of Electronic and Electrical Engineering, University College London, Adastral Park Postgraduate Research Campus, Martlesham Heath, IP5 3RE, United Kingdom (email: {l.choo, k.wong}@adastral.ucl.ac.uk).

In this paper, we investigate the degraded $K$-receiver BC with confidential messages sent to each receiver in the presence of a wiretapper, from which the messages are kept secret. Our results are a generalization of our work for 3 receivers in [11] and earlier results for 2 receivers in [10]. It is noted that results similar to ours have been established independently in [12], where Ekrem and Ulukus [12] examined the $K$-receiver degraded BC and one wiretapper with confidential messages as well as a common message sent to the receivers. However, there are some appreciable differences between our approach and that in [12]. In particular, equivocation calculation and proof of the converse in [12] are accomplished from the perspective of the channel sum rate. In contrast, we provide the error probability analysis and the equivocation calculation with respect to the $k$th receiver's achievable rate. Further, we use Wyner's method of random code partitioning instead of Gel'fand-Pinsker binning which is used in [10] to perform code partitioning to achieve perfect secrecy. In our proof of the converse, which we have shown for the $k$th receiver, we note that our choice of auxiliary random variables is different from that of [10] and [12]. Due to the presence of the wiretapper, it is also different from the choice in Borade *et al.* [13] where the capacity region for the degraded $K$-receiver BC using superposition coding without confidential messages is studied.

The remainder of this paper is organized as follows. In Section II, we introduce the general $K$-receiver degraded BC with confidential messages. In Section III, we state our main result for the secrecy capacity for the degraded $K$-receiver BC with wiretapper and show the proof of achievability and equivocation calculation in Sections III-A and III-B, respectively. In Section IV, we show the converse proof. Lastly we give conclusions in Section V.

## II. Channel model

In this paper, we use the uppercase letter to denote a random variable (e.g., $X$) and the lowercase letter for its realization (e.g., $x$). The alphabet set of $X$ is denoted by $\mathcal{X}$ so that $x \in \mathcal{X}$. We can also have a sequence of $n$ random variables, denoted by $\mathbf{X} = (X_1, \ldots, X_n)$ with its realization $\mathbf{x} = (x_1, \ldots, x_n) \in \mathcal{X}^n$ if $x_i \in \mathcal{X}$ for $i = 1, 2, \ldots, n$. Furthermore, we define the subsequences of $\mathbf{X}$ as $\mathbf{X}^i \triangleq (X_1, X_2, \ldots, X_i)$ and $\tilde{\mathbf{X}}^i \triangleq (X_i, \ldots, X_n)$.

The discrete memoryless $K$-receiver BC with an external wiretapper consists of a finite input alphabet $\mathcal{X}$ and finite output alphabets $\mathcal{Y}_1, \ldots, \mathcal{Y}_K, \mathcal{Z}$ and has conditional distribution $p(y_1, \ldots, y_K, z|x)$. Thus the discrete memoryless BC with $K$ receivers and a wiretapper has an input random sequence $\mathbf{X}$, $K$ output random sequences, $\mathbf{Y}_1, \ldots, \mathbf{Y}_K$, at the intended receivers, and an output random sequence at the wiretapper $\mathbf{Z}$. Likewise, we have $\mathbf{y}_1 \in \mathcal{Y}_1^n$, ..., $\mathbf{y}_K \in \mathcal{Y}_K^n$ and $\mathbf{z} \in \mathcal{Z}^n$. The conditional distribution for $n$ uses of the channel is

$$p(\mathbf{y}_1, \ldots, \mathbf{y}_K, \mathbf{z}|\mathbf{x}) = \prod_{i=1}^{n} p(y_{1i}, \ldots, y_{Ki}, z_i|x_i). \tag{1}$$

The transmitter has to send independent messages $(W_1, \ldots, W_K)$ to the receivers in perfect secrecy. This is done using a $(2^{nR_1}, \ldots, 2^{nR_K}, n)$-code for the BC, which consists of the stochastic encoder

$$f : \left\{1, \ldots, 2^{nR_1}\right\} \times \left\{1, \ldots, 2^{nR_2}\right\} \times \cdots \times \left\{1, \ldots, 2^{nR_K}\right\} \mapsto \mathcal{X}^n, \tag{2}$$

and the decoders

$$g_k : \mathcal{Y}_k^n \mapsto \left\{1, \ldots, 2^{nR_k}\right\}, \quad \text{for } k = 1, 2, \ldots, K. \tag{3}$$

The probability of error is defined as the probability that the decoded messages are not equal to the transmitted messages, i.e.,

$$P_e^{(n)} \triangleq \Pr\left\{\bigcup_{k=1}^{K} \{g_k(\mathbf{Y}_k) \neq W_k\}\right\}. \tag{4}$$

Perfect secrecy requires that the mutual information of the transmitted messages and the wiretapper goes to zero. Let us illustrate this for message $W_1$ and receiver 1. The perfect secrecy requirement is

$$I(W_1;\mathbf{Z}) = 0 \quad \Rightarrow \quad H(W_1) = H(W_1|\mathbf{Z}), \tag{5}$$

where $I(\cdot;\cdot)$ denotes mutual information and $H(\cdot)$ is entropy. Now, let the information rate for the first receiver be $R_1 = \frac{1}{n}H(W_1)$ and the equivocation rate be $R_{e(1)} \triangleq \frac{1}{n}H(W_1|\mathbf{Z})$. Then, we need

$$R_{e(1)} \geq R_1 - \eta, \quad \text{for any arbitrary } \eta > 0. \tag{6}$$

Further to this, we define the following equivocation rates for the $K$-receiver degraded BC:

$$\begin{cases} R_{e(k)} \triangleq \dfrac{1}{n}H(W_k|\mathbf{Z}), & \text{for } k = 1, \ldots, K, \\ R_{e(k,k+1)} \triangleq \dfrac{1}{n}H(W_k, W_{k+1}|\mathbf{Z}), & \text{for } k = 1, \ldots, K, \\ R_{e(1,\ldots,K)} \triangleq \dfrac{1}{n}H(W_1, \ldots, W_K|\mathbf{Z}). \end{cases} \tag{7}$$

## III. The Secrecy Capacity Region

The secret rate tuple $(R_1, R_2, \ldots, R_K)$ is achievable if for any arbitrarily small $\epsilon' > 0$, $\epsilon_k > 0$, $k = 1, \ldots, K$, $\epsilon_{k,k+1} > 0$, $k = 1, 2, \ldots, K-1$, and $\epsilon_{1,\ldots,K} > 0$, there exist $(2^{nR_1}, \ldots, 2^{nR_K}, n)$-codes for which $P_e^{(n)} \leq \epsilon'$ and

$$\begin{cases} R_{e(k)} \geq R_k - \epsilon_k, & \text{for } k = 1, \ldots, K, \\ R_{e(k,k+1)} \geq R_k + R_{k+1} - \epsilon_{k,k+1}, & \text{for } k = 1, \ldots, K, \\ R_{e(1,\ldots,K)} \geq \displaystyle\sum_{k=1}^{K} R_k - \epsilon_{1,\ldots,K}. \end{cases} \tag{8}$$

(8) gives the security conditions for the $K$-receiver BC with a wiretapper under perfect secrecy requirements in (6).

*Theorem 1:* The secrecy capacity region for the $K$-receiver degraded BC with an external wiretapper is the closure of all rate tuples $(R_1, \ldots, R_K)$ satisfying

$$R_1 \leq I(X;Y_1|U_2) - I(X;Z|U_2), \tag{9a}$$

$$R_k \leq I(U_k;Y_k|U_{k+1}) - I(U_k;Z|U_{k+1}), \quad \text{for } k = 2, \ldots, K-1, \tag{9b}$$

$$R_K \leq I(U_K;Y_K) - I(U_K;Z), \tag{9c}$$

where $\{U_k\}_{k=2}^{K}$ are auxiliary random variables and will be defined in Section III-A (Random codebook generation).

*Proof:* The proof of achievability and equivocation calculation are given later in this section. The proof of converse is given separately in Section IV. ■

If we use superposition coding with code partitioning to achieve the rates in Theorem 1, then the secrecy capacity region may be interpreted as the capacity region for the $K$-receiver BC using superposition coding without the wiretapper, with the rates at each receiver each reduced due to the presence of the wiretapper. However, we shall see that the choice of auxiliary random variables in the proof of converse for the $K$-receiver BC will be different from that of [13], which is without the secrecy conditions. This is also in contrast to the 2-receiver BC with wiretapper in [10], where the same definition for the auxiliary random variables in the converse proof can be used for the scenarios with and without the secrecy conditions.

### *A. Proof of Achievability*

In this paper, we employ superposition coding and Wyner's random code partitioning to show the achievable rate tuples $(R_1, \ldots, R_K)$. For brevity, we use $p_{\mathbf{Y}_1|\mathbf{X}}$ to denote the channel from $\mathbf{X}$ to $\mathbf{Y}_1$, similarly for the channels from $\mathbf{X}$ to outputs $\mathbf{Y}_2$, ...,$\mathbf{Y}_K$ and $\mathbf{Z}$, by $p_{\mathbf{Y}_2|\mathbf{X}}, \ldots, p_{\mathbf{Y}_K|\mathbf{X}}$ and $p_{\mathbf{Z}|\mathbf{X}}$, respectively.

The coding strategy is depicted in Fig. 1. The message $W_k \in \{1, \ldots, L_k\}$ with $L_k \triangleq 2^{nR_k}$ for $k = 1, \ldots, K$, is sent by a code of length $N_k = L_k L'_k$. This code is partitioned into $L_k$ subcodes each of size $L'_k$, with $L'_k \triangleq 2^{nR'_k}$ for some $R'_k$. Each of the $L_k$ subcodes is a code for the wiretapper $p_{\mathbf{Z}|\mathbf{X}}$, while each of the entire codes of size $N_k$ is a code simultaneously for both the $k$th receiver $p_{\mathbf{Y}_k|\mathbf{X}}$ and the wiretapper $p_{\mathbf{Z}|\mathbf{X}}$. The codes for simultaneous use for $p_{\mathbf{Y}_k|\mathbf{X}}$ and $p_{\mathbf{Z}|\mathbf{X}}$ have to satisfy the transmission requirements for the BC [14], so that

$$\frac{1}{n}\log N_1 \le I(X;Y_1|U_2), \tag{10a}$$

$$\frac{1}{n}\log N_k \le I(U_k;Y_k|U_{k+1}), \quad \text{for } k = 2, \ldots, K-1, \tag{10b}$$

$$\frac{1}{n}\log N_K \le I(U_K;Y_K). \tag{10c}$$

1) Random codebook generation:

Suppose that we have the probability density functions (p.d.f.s)

$$\begin{cases} p(u_K), \\ p(u_k|u_{k+1}, \ldots, u_K), \quad \text{for } k = 2, \ldots, K-1, \\ p(x|u_2, \ldots, u_K). \end{cases} \tag{11}$$

For a given rate tuple $(R_1, \ldots, R_K, R'_1, \ldots, R'_K)$, in order to send message $W_K$, generate $2^{n(R_K+R'_K)}$ independent codewords $\mathbf{u}_K(w''_K)$, for $w''_K \in \{1, \ldots, 2^{n(R_K+R'_K)}\}$ according to the p.d.f. $p(\mathbf{u}_K) = \prod_{i=1}^n p(u_{Ki})$. Then, partition $\mathbf{u}_K(w''_K)$ into $L_K = 2^{nR_K}$ subcodes, $\{\mathcal{C}_i^{(K)}\}_{i=1}^{L_K}$ with $|\mathcal{C}_i^{(K)}| = L'_K = 2^{nR'_K}$ $\forall i$.

The message for the $k$th receiver, for $k = 2, 3, \ldots, K-1$, is sent by generating $2^{n(R_k+R'_k)}$ independent codewords $\mathbf{u}_k(w''_k, \ldots, w''_K)$, for $w''_k \in \{1, \ldots, 2^{n(R_k+R'_k)}\}$ according to the conditional p.d.f.

$$p(\mathbf{u}_k|\mathbf{u}_{k+1}, \ldots, \mathbf{u}_K) = \prod_{i=1}^n p(u_{ki}|u_{(k+1)i}, \ldots, u_{Ki}). \tag{12}$$

Then, partition $\mathbf{u}_k(w''_k, \ldots, w''_K)$ into $L_k = 2^{nR_k}$ subcodes, $\{\mathcal{C}_i^{(k)}\}_{i=1}^{L_k}$, with $|\mathcal{C}_i^{(k)}| = L'_k = 2^{nR'_k}$ $\forall i$. Finally, to send the message intended for the first receiver, generate $2^{n(R_1+R'_1)}$ independent codewords $\mathbf{x}(w''_1, \ldots, w''_K)$, for $w''_1 \in \{1, \ldots, 2^{n(R_1+R'_1)}\}$ according to the p.d.f. $p(\mathbf{x}|\mathbf{u}_2, \ldots, \mathbf{u}_K) = \prod_{i=1}^n p(x|u_{2i}, \ldots, u_{Ki})$. Then, partition $\mathbf{x}(w''_1, \ldots, w''_K)$ into $L_1 = 2^{nR_1}$ subcodes, $\{\mathcal{C}_i^{(1)}\}_{i=1}^{L_1}$, with $|\mathcal{C}_i^{(1)}| = L'_1 = 2^{nR'_1}$ $\forall i$.

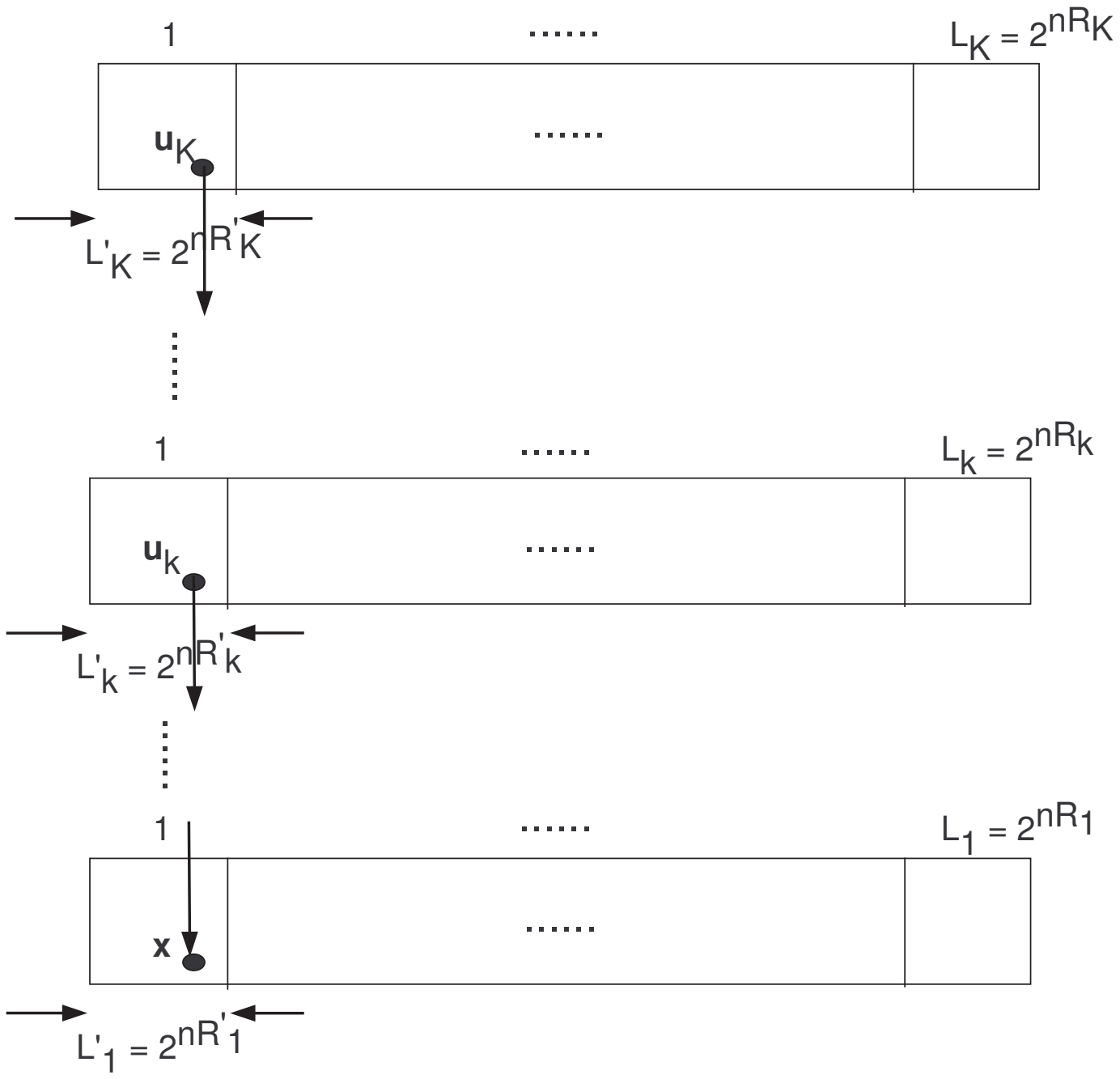

Fig. 1. Coding for $K$ receiver BC with wiretapper.

Following this code structure, the codeword indices $w''_k$ may be expressed as $w''_k = (w_k, w'_k)$, where $w_k \in \{1, \ldots, 2^{nR_k}\}$ is the index of the message transmitted to the $k$th receiver, and $w'_k \in \{1, \ldots, 2^{nR'_k}\}$ denotes the index of the codeword within the subcodes $\mathcal{C}_i^{(k)}$, selected for transmission along with $w_k$ to ensure secrecy.

2) Encoding:

The encoding is by superposition coding. To send the message $w_K = i_K$, for $1 \leq i_K \leq L_K$, the transmitter chooses one of the $\mathbf{u}_K(w''_K)$ codewords uniformly and randomly from $\{\mathcal{C}_{i_K}^{(K)}\}_{i_K=1}^{L_K}$. Then, to send the message $w_{K-1} = i_{K-1}$, for $1 \leq i_{K-1} \leq L_{K-1}$, the transmitter selects one of the $\mathbf{u}_{K-1}(w''_{K-1}, w''_K)$ uniformly randomly from $\{\mathcal{C}_{i_{K-1}}^{(K-1)}\}_{i_{K-1}=1}^{L_{K-1}}$, given $\mathbf{u}_K(w''_K)$. Sequentially, the transmitter sends the message $w_k = i_k$, for $1 \leq i_k \leq L_k$ and $k = 2, \ldots, K-2$, to the $k$th receiver by choosing one of the $\mathbf{u}_k(w''_k, \ldots, w''_K)$ uniformly and randomly from $\{\mathcal{C}_{i_k}^{(k)}\}_{i_k=1}^{L_k}$, given $\mathbf{u}_{k+1}(w''_{k+1}, \ldots, w''_K)$. Lastly, to send $w_1 = i_1$ for $1 \leq i_1 \leq L_1$, given $\mathbf{u}_2(w''_2, \ldots, w''_K)$, the transmitter chooses one of the $\mathbf{x}(w''_1, \ldots, w''_K)$ uniformly randomly from $\{\mathcal{C}_{i_1}^{(1)}\}_{i_1=1}^{L_1}$.

3) Decoding:

We use the notation $A_\epsilon^n(P_V)$ to denote the set of jointly typical $n$-sequences with respect to the p.d.f. $p(\mathbf{v})$. Also, we use $\{\hat{w}_k\}_{k=1}^K$ to denote the estimates for the transmitted messages $\{w_k\}_{k=1}^K$. Then, we have:

a) At the $K$th receiver, given that $\mathbf{y}_K$ is received, find a $\hat{w}_K$, such that $(\mathbf{u}_K(\hat{w}_K, w'_K), \mathbf{y}_K) \in A_\epsilon^n(P_{U_K Y_K})$.

b) At the $k$th receiver, for $k = 2, \ldots, K-1$, given that $\mathbf{y}_k$ is received, find a $(\hat{w}_k, \ldots, \hat{w}_K)$ such that

$$(\mathbf{u}_K(\hat{w}_k, w'_k), \ldots, \mathbf{u}_k(\hat{w}_k, w'_k, \ldots, \hat{w}_K, w'_K), \mathbf{y}_k) \in A_\epsilon^n(P_{U_K U_{K-1} \cdots U_k Y_k}). \tag{13}$$

c) Lastly, at the first receiver, given that $\mathbf{y}_1$ is received, find a $(\hat{w}_1, \ldots, \hat{w}_K)$ such that

$$(\mathbf{u}_K(\hat{w}_K, w'_K), \ldots, \mathbf{u}_2(\hat{w}_2, w'_2, \ldots, \hat{w}_K, w'_K), \mathbf{x}_1(\hat{w}_1, w'_1, \ldots, \hat{w}_K, w'_K), \mathbf{y}_1) \in A^n_\epsilon(P_{U_K U_{K-1} \cdots U_2 X Y_1}). \tag{14}$$

For each of the above cases, if there is none or more than one possible decoded message, then an error will be declared. Note that $w'_k$ is unimportant for the decoding of $w_k$ at the $k$th receiver.

4) Obtaining the sizes of subcodes $\{\mathcal{C}^{(k)}_{i_k}\}$:

Here, we shall not use binning but follow the approach of Wyner [7], where random code partitioning is used. We shall show how to obtain $\log L'_k$ in the encoding of $W_k$, for $k = 2, ..., K-1$. Following the same routine, $\log L'_1$ and $\log L'_K$ can be obtained easily, and thus these calculations will be omitted.

To start with, suppose that we have the messages, $w_k = i_k, \ldots, w_K = i_K$. We now define

$$\begin{aligned} q^{(k)}_{i_k} &\triangleq \Pr\{W_k = i_k | W_{k+1} = i_{k+1}, \ldots, W_K = i_K\} \\ &= \Pr\left\{W_k = i_k | \mathbf{u}_K(i_K, i'_K), \mathbf{u}_{K-1}(i_{K-1}, i'_{K-1}, i_K, i'_K), \ldots, \mathbf{u}_k(i_k, i'_k, \ldots, i_K, i'_K)\right\}. \end{aligned} \tag{15}$$

The codeword $\mathbf{u}_k(w''_k, \ldots, w''_K)$ is a channel code for $p_{\mathbf{Y}_k|\mathbf{X}}$ and $p_{\mathbf{Z}|\mathbf{X}}$ simultaneously and is comprised of $L_k = 2^{nR_k}$ subcodes $\{C^{(k)}_{i_k}\}^{L_k}_{i_k=1}$. $\mathbf{U}_k$ is an uniformly randomly chosen member of $\{C^{(k)}_{i_k}\}$. Therefore,

$$\Pr\left\{\mathbf{U}_k = \mathbf{u}_k(w''_k, \ldots, w''_K) | \mathbf{u}_K(i_K, i'_K), \ldots, \mathbf{u}_{k+1}(i_{k+1}, i'_{k+1}, \ldots, i_K, i'_K)\right\} = \frac{q^{(k)}_{i_k}}{L'_k}. \tag{16}$$

The codeword $\mathbf{u}_k(w''_k, \ldots, w''_K)$ is a channel code for $p_{\mathbf{Y}_k|\mathbf{X}}$ with prior distribution on codewords given by (16). Each of $\mathcal{C}^{(k)}_{i_k}$ is a channel code for the wiretap channel $p_{\mathbf{Z}|\mathbf{X}}$ with $L'_k$ codewords and uniform prior distribution on the codewords. Let $\lambda^{(k)}_{i_k}$ be the error probability for $\mathcal{C}^{(k)}_{i_k}$ with an optimal decoder, when $i'_k$ is chosen as the index for the codeword from $\mathcal{C}^{(k)}_{i_k}$. Then $\bar{\lambda}^{(k)}$ is the average error probability for $\mathcal{C}^{(k)}_{i_k}$ with an optimal decoder, averaged over the probability that $W_k = i_k$ is sent given the previous messages were $W_{k+1} = i_{k+1}, \ldots W_K = i_K$. As a result, we have

$$\begin{cases} \lambda^{(k)}_{i_k} = \Pr\left\{\mathbf{X} \neq \mathbf{Z} | W_k = i_k, \mathbf{u}_K(i_K, i'_K), \ldots, \mathbf{u}_{k+1}(i_{k+1}, i'_{k+1}, \ldots, i_K, i'_K)\right\}, \\ \bar{\lambda}^{(k)} = \sum_{i_k=1}^{L_k} q^{(k)}_{i_k} \lambda^{(k)}_{i_k}. \end{cases} \tag{17}$$

By Fano's inequality,

$$\begin{aligned} H(\mathbf{X}|\mathbf{Z}, W_k = i_k, \mathbf{u}_K(i_K, i'_K), \ldots, \mathbf{u}_{k+1}(i_{k+1}, i'_{k+1}, \ldots, i_K, i'_K)) &\leq 1 + \lambda^{(k)}_{i_k} \log L'_k \\ \Rightarrow \quad H(\mathbf{U}_k|\mathbf{Z}, \mathbf{U}_K, \ldots, \mathbf{U}_{k+1}, W_k = i_k) &\leq 1 + \lambda^{(k)}_{i_k} \log L'_k. \end{aligned} \tag{18}$$

Since $|\mathcal{C}^{(k)}_{i_k}| = L'_k$ and has probability of error $\lambda^{(k)}_{i_k}$, we have

$$\begin{aligned} I(\mathbf{U}_k; \mathbf{Z}|\mathbf{U}_K, \ldots, \mathbf{U}_{k+1}, W_k = i_k) &= H(\mathbf{U}_k|\mathbf{U}_K, \ldots, \mathbf{U}_{k+1}, W_k = i_k) - H(\mathbf{U}_k|\mathbf{Z}, \mathbf{U}_K, \ldots, \mathbf{U}_{k+1}, W_k = i_k) \\ &= \log L'_k - H(\mathbf{U}_k|\mathbf{Z}, \mathbf{U}_K, \ldots, \mathbf{U}_{k+1}, W_k = i_k) \\ \Rightarrow \quad \log L'_k &\leq I(\mathbf{U}_k; \mathbf{Z}|\mathbf{U}_K, \ldots, \mathbf{U}_{k+1}, W_k = i_k) + 1 + \lambda^{(k)}_{i_k} \log L'_k. \end{aligned} \tag{19}$$

Averaging over $i_k$ using $\{q_{i_k}^{(k)}\}$ gives

$$\begin{aligned}\log L'_k &\le I(\mathbf{U}_k;\mathbf{Z}|\mathbf{U}_K,\ldots,\mathbf{U}_{k+1},W_k)+1+\bar{\lambda}^{(k)}\log L'_k\\ &\overset{(a)}{\le} I(\mathbf{U}_k;\mathbf{Z}|\mathbf{U}_K,\ldots,\mathbf{U}_{k+1})+1+\bar{\lambda}^{(k)}\log L'_k\\ &\overset{(b)}{\le} nI(U_k;Z|U_K,\ldots,U_{k+1})+n\delta+1+\bar{\lambda}^{(k)}\log L'_k,\\ &\overset{(c)}{=} nI(U_k;Z|U_{k+1})+n\delta+1+\bar{\lambda}^{(k)}\log L'_k,\end{aligned} \tag{20}$$

where (a) is by $W_k \to (\mathbf{U}_K,\ldots,\mathbf{U}_{k+1}) \to \mathbf{U}_k \to \mathbf{Z}$, (b) results from the fact that (following Liu *et al.* [8])

$$I(\mathbf{U}_k;\mathbf{Z}|\mathbf{U}_K,\ldots,\mathbf{U}_{k+1}) \le nI(U_k;Z|U_K,\ldots,U_{k+1})+n\delta, \tag{21}$$

with $\delta \to 0$ as $n \to \infty$ and (c) is by the Markov chain condition $U_K \to \cdots \to U_{k+1} \to U_k \to Z$ for the degraded BC. Similarly, by substituting $\mathbf{X}$ for $\mathbf{U}_1$ and removing conditioning from (15) for $k=K$, we have

$$\begin{cases}\log L_1 \le nI(X;Z|U_2)+n\delta+1+\bar{\lambda}^{(1)}\log L'_1,\\ \log L_K \le nI(U_K;Z)+n\delta+1+\bar{\lambda}^{(K)}\log L'_K.\end{cases} \tag{22}$$

Based on the above, and since $R'_k=\frac{1}{n}\log L'_k$, we let

$$\begin{cases}R'_1 \triangleq I(X;Z|U_2)-\tau,\\ R'_k \triangleq I(U_k;Z|U_{k+1})-\tau, \quad \text{for } k=2,\ldots,K-1,\\ R'_K \triangleq I(U_K;Z)-\tau,\end{cases} \tag{23}$$

where $\tau \to 0$ for sufficiently large $n$.

5) Probability of error analysis:

We follow the method by Cover and Thomas in [15], and provide the analysis for the $k$th receiver. Assume without loss of generality that $(W_1,\ldots,W_k)=(1,\ldots,1)$ is sent and $w'_k=1$ is sent for the subcodes $\mathcal{C}_{i_k}^{(k)}$ $\forall k$. At the $k$th receiver, define the following events (and their complements denoted by the superscript c):

$$\begin{cases}\mathrm{E}_{i_K,i'_K}^{(Y_k)} = \left\{(\mathbf{U}_K(i_K,i'_K),\mathbf{Y}_k)\in A_\epsilon^{(n)}\right\}\\ \mathrm{E}_{i_K,i'_K,i_{K-1},i'_{K-1}}^{(Y_k)} = \left\{(\mathbf{U}_K(i_K,i'_K),\mathbf{U}_{K-1}(i_K,i'_K,i_{K-1},i'_{K-1}),\mathbf{Y}_k)\in A_\epsilon^{(n)}\right\}\\ \vdots\\ \mathrm{E}_{i_K,i'_K,\ldots,i_k,i'_k}^{(Y_k)} = \left\{(\mathbf{U}_K(i_K,i'_K),\ldots,\mathbf{U}_k(i_K,i'_K,\ldots,i_k,i'_k),\mathbf{Y}_k)\in A_\epsilon^{(n)}\right\}.\end{cases} \tag{24}$$

Then, by the union of events bound, we have

$$\begin{aligned}P_e^{(n)}(k) \le{}& \Pr\left\{\left(\mathrm{E}_{1,1}^{(Y_k)}\right)^{\mathrm{c}}\right\}+\Pr\left\{\left(\mathrm{E}_{1,1,1,1}^{(Y_k)}\right)^{\mathrm{c}}\right\}+\cdots+\Pr\left\{\left(\mathrm{E}_{1,1,\ldots,1,1}^{(Y_k)}\right)^{\mathrm{c}}\right\}+\\ &\sum_{\substack{i_K,i'_K\\ i_K\neq 1}}\Pr\left\{\left(\mathrm{E}_{i_K,i'_K}^{(Y_k)}\right)^{\mathrm{c}}\right\}+\sum_{\substack{i_{K-1},i'_{K-1}\\ i_{K-1}\neq 1}}\Pr\left\{\left(\mathrm{E}_{1,1,i_{K-1},i'_{K-1}}^{(Y_k)}\right)^{\mathrm{c}}\right\}+\cdots+\sum_{\substack{i_k,i'_k\\ i_k\neq 1}}\Pr\left\{\left(\mathrm{E}_{1,1,\ldots,i_k,i'_k}^{(Y_k)}\right)^{\mathrm{c}}\right\},\end{aligned} \tag{25}$$

where there are $(K-k+1)$ terms in each of the first and second lines of the inequality (25) above; the last term in the first line of (25) refers to the probability that the complement of the event that the $2k$-length vector of all ones $(1,1,\ldots,1,1)$ occurred; and the last term in the second line of (25) refers to the probability that the event that the $2k$-length vector $(1,1,\ldots,i_k,i'_k)$ occurred. For the first two terms of (25), we have

$$\begin{cases} \Pr\left\{\mathtt{E}^{(Y_1)}_{i_K,i'_K}\right\} \le 2^{-n(I(U_K;Y_k)-3\epsilon)}, \\ \Pr\left\{\mathtt{E}^{(Y_1)}_{1,1,i_{K-1},i'_{K-1}}\right\} \le 2^{-n(I(U_{K-1};Y_k|U_K)-4\epsilon)}. \end{cases} \tag{26}$$

Denoting the event $\{(\mathbf{U}_K,\ldots,\mathbf{U}_k,\mathbf{Y}_k)\in A_\epsilon^{(n)}\}$ as $\tilde{\mathtt{E}}^{(Y_k)}$, the $k$th term can be written as

$$\begin{aligned}
&\Pr\left\{\left(\mathtt{E}^{(Y_k)}_{1,1,\ldots,i_k,i'_k}\right)^{\mathsf{c}}\right\} = \Pr\left\{(\mathbf{U}_K(1,1),\mathbf{U}_{K-1}(1,1,1,1),\ldots,\mathbf{U}_k(1,1,\ldots,i_k,i'_k),\mathbf{Y}_k)\in A_\epsilon^{(n)}\right\} \\
&= \sum_{\tilde{\mathtt{E}}^{(Y_k)}} \Pr\left\{(\mathbf{U}_K(1,1),\mathbf{U}_{K-1}(1,1,1,1),\ldots,\mathbf{U}_k(1,1,\ldots,i_k,i'_k),\mathbf{Y}_k)\right\} \\
&= \sum_{\tilde{\mathtt{E}}^{(Y_k)}} \left[\begin{array}{l} \Pr\{\mathbf{U}_K(1,1)\}\Pr\{\mathbf{U}_{K-1}(1,1,1,1)|\mathbf{U}_K(1,1)\}\times\cdots\times \\ \Pr\left\{\mathbf{U}_k(1,1,\ldots,i_k,i'_k)|\mathbf{U}_K(1,1),\mathbf{U}_{K-1}(1,1,1,1),\ldots,\mathbf{U}_{k+1}(\overbrace{1,1,\ldots,1,1}^{2(k+1)\text{ terms}})\right\}\times \\ \Pr\left\{\mathbf{Y}_k|\mathbf{U}_K(1,1),\mathbf{U}_{K-1}(1,1,1,1),\ldots,\mathbf{U}_{k+1}(\overbrace{1,1,\ldots,1,1}^{2(k+1)\text{ terms}})\right\} \end{array}\right] \\
&\le \sum_{\tilde{\mathtt{E}}^{(Y_k)}} 2^{-n(H(U_K)-\epsilon)}2^{-n(H(U_{K-1}|U_K)-\epsilon)}\times\cdots\times 2^{-n(H(U_k|U_K,\ldots,U_{k+1})-\epsilon)}2^{-n(H(Y_k|U_K,\ldots,U_{k+1})-\epsilon)} \\
&\le 2^{n(H(U_K,\ldots,U_{k+1},U_k,Y_k)+\epsilon)}2^{-n(H(U_K)-\epsilon)}2^{-n(H(U_{K-1}|U_K)-\epsilon)} \\
&\quad \times\cdots\times 2^{-n(H(U_k|U_K,\ldots,U_{k+1})-\epsilon)}2^{-n(H(Y_k|U_K,\ldots,U_{k+1})-\epsilon)} \\
&= 2^{-n(H(Y_k|U_K,\ldots,U_{k+1})-H(Y_k|U_K,\ldots,U_{k+1},U_k)-(k+2)\epsilon)} = 2^{-n(I(Y_k;U_k|U_K,\ldots,U_{k+1})-(k+2)\epsilon)}.
\end{aligned}$$

As a result,

$$\begin{aligned}
P_e^{(n)}(k) &\le (K-k+1)\epsilon + 2^{n(R_K+R'_K)}2^{-n(I(U_K;Y_k)-3\epsilon)} + 2^{n(R_{K-1}+R'_{K-1})}2^{-n(I(U_{K-1};Y_k|U_K)-4\epsilon)} \\
&+\cdots+2^{n(R_k+R'_k)}2^{-n(I(Y_k;U_k|U_K,\ldots,U_{k+1})-(k+2)\epsilon)} \le 2(K-k+1)\epsilon, \quad \text{for } n \text{ sufficiently large and}
\end{aligned} \tag{27}$$

$$R_K + R'_K < I(U_K;Y_k), \tag{28a}$$

$$R_{K-1}+R'_{K-1} < I(U_{K-1};Y_k|U_K), \tag{28b}$$

$$\vdots$$

$$R_k + R'_k < I(U_k;Y_k|U_K,\ldots,U_{k+1}). \tag{28c}$$

Since $I(U_K;Y_k) \ge I(U_K;Y_{k+1}) \ge \cdots \ge I(U_K;Y_K)$ and similarly $I(U_{k+1};Y_k|U_K,\ldots,U_{k+2}) \ge \cdots \ge$

$I(U_{k+1};Y_{k+1}|U_K,\dots,U_{k+2})$ by the degraded nature of the channel, from (28a), we get

$$R_K + R'_K \leq I(U_K;Y_K), \tag{29a}$$

$$R_{K-1} + R'_{K-1} \leq I(U_{K-1};Y_{K-1}|U_K), \tag{29b}$$

$$\vdots$$

$$R_k + R'_k \leq I(U_k;Y_k|U_K,\dots U_{k+1}), \tag{29c}$$

for the second to the last terms in (27) to be $\leq \epsilon$. Then, as we have the condition $U_K \to U_{K-1} \to \cdots \to U_{k+1} \to U_k \to Y_k$, we have

$$R_K + R'_K \leq I(U_K;Y_K), \tag{30a}$$

$$R_{K-1} + R'_{K-1} \leq I(U_{K-1};Y_{K-1}|U_K), \tag{30b}$$

$$\vdots$$

$$R_k + R'_k \leq I(U_k;Y_k|U_{k+1}). \tag{30c}$$

Following the same approach, for the first receiver, we can get the above inequalities in (30), as well as

$$R_1 + R'_1 \leq I(X;Y_1|U_2). \tag{31}$$

Therefore, for all the receivers, given the previous definitions for $R'_1,\dots,R'_K$ in (23), it can be seen that the probability of error at each receiver satisfies $P_e^{(n)}(k) \leq 2(K-k+1)\epsilon$ for $k=1,\dots,K$ and for any rate tuple $(R_1,\dots,R_K)$ satisfying the conditions in Theorem 1.

Thus, the direct part of Theorem 1 is proved.

### B. *Equivocation Calculation*

We show the calculation for the $k$th receiver $R_{e(k)}$ for $k=1,\dots,K$, $R_{e(k,k+1)}$ for $k=1,\dots,K-1$, and the sum rate $R_{e(1,\dots,K)}$. We shall make use of the relation

$$H(U,V) = H(U) + H(V|U). \tag{32}$$

For the $k$th receiver, $k=2,\dots,K-1$, we have

$$\begin{aligned}
nR_{e(k)} &= H(W_k|\mathbf{Z}) \\
&\geq H(W_k|\mathbf{Z},\mathbf{U}_K,\dots,\mathbf{U}_{k+1}) \quad \text{since conditioning reduces entropy} \\
&= H(W_k,\mathbf{Z}|\mathbf{U}_K,\dots,\mathbf{U}_{k+1}) - H(\mathbf{Z}|\mathbf{U}_K,\dots,\mathbf{U}_{k+1}) \quad \text{by (32)} \\
&\overset{(a)}{=} H(W_k,\mathbf{U}_k,\mathbf{Z}|\mathbf{U}_K,\dots,\mathbf{U}_{k+1}) - H(\mathbf{U}_k|W_k,\mathbf{Z},\mathbf{U}_K,\dots,\mathbf{U}_{k+1}) - H(\mathbf{Z}|\mathbf{U}_K,\dots,\mathbf{U}_{k+1}) \\
&\overset{(b)}{=} H(W_k,\mathbf{U}_k|\mathbf{U}_K,\dots,\mathbf{U}_{k+1}) + H(\mathbf{Z}|W_k,\mathbf{U}_K,\dots,\mathbf{U}_{k+1},\mathbf{U}_k) - H(\mathbf{Z}|\mathbf{U}_K,\dots,\mathbf{U}_{k+1}) \\
&\quad - H(\mathbf{U}_k|W_k,\mathbf{Z},\mathbf{U}_K,\dots,\mathbf{U}_{k+1})
\end{aligned}$$

$$\overset{(c)}{\geq} H(\mathbf{U}_k|\mathbf{U}_K,\ldots,\mathbf{U}_{k+1}) + H(\mathbf{Z}|\mathbf{U}_K,\ldots,\mathbf{U}_{k+1},\mathbf{U}_k) - H(\mathbf{Z}|\mathbf{U}_K,\ldots,\mathbf{U}_{k+1})$$
$$- H(\mathbf{U}_k|W_k,\mathbf{Z},\mathbf{U}_K,\ldots,\mathbf{U}_{k+1})$$
$$= H(\mathbf{U}_k|\mathbf{U}_K,\ldots,\mathbf{U}_{k+1}) - I(\mathbf{U}_k;\mathbf{Z}|\mathbf{U}_K,\ldots,\mathbf{U}_{k+1}) - H(\mathbf{U}_k|W_k,\mathbf{Z},\mathbf{U}_K,\ldots,\mathbf{U}_{k+1}), \tag{33}$$

where (a) and (b) have the first two terms by (32), and (c) has the first term by (32) and the second term by the fact that $W_k \to (\mathbf{U}_K,\ldots,\mathbf{U}_{k+1}) \to \mathbf{Z}$. We now bound each of the terms in the last line of (33). For the first term, given that $\mathbf{U}_K = \mathbf{u}_K$, $\mathbf{U}_{K-1} = \mathbf{u}_{K-1},\ldots,\mathbf{U}_{k+1} = \mathbf{u}_{k+1}$, $\mathbf{u}_k$ has $2^{n(R_k+R'_k)}$ possible values with equal probability. As a consequence, we have

$$H(\mathbf{U}_k|\mathbf{U}_K,\ldots,\mathbf{U}_{k+1}) = n(R_k + R'_k). \tag{34}$$

For the second term, it can be shown that

$$I(\mathbf{U}_k;\mathbf{Z}|\mathbf{U}_K,\ldots,\mathbf{U}_{k+1}) \leq nI(U_k;Z|U_{k+1}) + n\delta. \tag{35}$$

For the last term, we have by Fano's inequality

$$\frac{1}{n}H(\mathbf{U}_k|W_k,\mathbf{Z},\mathbf{U}_K,\ldots,\mathbf{U}_{k+1}) \leq \frac{1}{n}\left(1 + \bar{\lambda}^{(k)}\log L'_k\right) \triangleq \epsilon'_{k,n} \tag{36}$$

where $\epsilon'_{k,n} \to 0$ for $n$ sufficiently large.

To show that $\bar{\lambda}^{(k)} \to 0$ for $n$ sufficiently large so that (36) holds, we consider decoding at the wiretapper and focus on the codebook with rate $R'_k$ to be decoded at the wiretapper with error probability $\bar{\lambda}^{(k)}$. Let $W_k = i_k$ be fixed. The wiretapper attempts to decode $\mathbf{u}_k$ given $w_k, \mathbf{u}_K, \ldots, \mathbf{u}_{k+1}$ by finding the estimate for $w'_k$, $\hat{w}'_k$, so that

$$(\mathbf{u}_k(w_k,\hat{w}'_k,w_{k+1},w'_{k+1},\ldots,w_K,w'_K),\mathbf{z}) \in A^n_\epsilon(P_{U_kZ|U_{k+1}\ldots U_K}). \tag{37}$$

where $w_k$, and all $w_{k+1},w'_{k+1},\ldots,w_K,w'_K$ are known. If there is none or more than one possible codeword, an error is declared. Defining the event

$$\mathrm{E}^{(Z)}_{i'_k} \triangleq \left\{(\mathbf{U}_k(i_k,i'_k),\mathbf{Z}) \in A^{(n)}_\epsilon(P_{U_kZ|U_{k+1}\ldots U_K})\right\}, \tag{38}$$

and assuming without loss of generality that $w'_k = 1$ is sent, we then have

$$\bar{\lambda}^{(k)} \leq \Pr\left\{\left(\mathrm{E}^{(Z)}_1\right)^{\mathsf{c}}\right\} + \sum_{i'_k \neq 1}\Pr\left\{\left(\mathrm{E}^{(Z)}_{i'_k}\right)\right\} \leq \epsilon + 2^{nR'_k}2^{-n(I(U_k;Z|U_{k+1},\ldots,U_K)-2\epsilon)}, \tag{39}$$

where $\epsilon \to 0$ for $n$ sufficiently large. Since we have chosen from (23) that $R'_k = I(U_k|Z|U_{k+1}) - \tau$ which is $= I(U_k|Z|U_{k+1},\ldots,U_K) - \tau$ by $U_K \to \cdots \to U_{k+1} \to U_k \to Z$, we have $\bar{\lambda}^{(k)} \leq 2\epsilon$, for $\tau > 2\epsilon$. Thus, $\bar{\lambda}^{(k)}$ is small for $n$ sufficiently large and (36) holds.

Now substituting (34)–(36) into the last line of (33), we have

$$\begin{aligned} nR_{e(k)} &\geq nR_k + nI(U_k;Z|U_{k+1}) - n\tau - nI(U_k;Z|U_{k+1}) - n\delta - n\epsilon'_{k,n} \\ &= nR_k - n\epsilon_k \end{aligned} \tag{40}$$

where $\epsilon_k = \tau + \delta + \epsilon'_{k,n}$. Hence, the security condition in (8) is satisfied for the $k$th receiver. For the first receiver, we condition on $\mathbf{U}_K,\ldots,\mathbf{U}_2$ in the second line of the chain of inequalities above, while for the $K$th receiver, we can omit the second line of the chain of inequalities in (33) above, while subsequently not performing additional conditioning in (33). The equivocation rates for the first receiver and the $K$th receiver will have the same form as (40) above with $k = 1$ and $k = K$.

We next show the equivocation rates for adjacent receivers $k, k+1$ for $k = 1, \ldots, K-1$. Due to the nature of the coding, equivocation rates for non-adjacent receivers are not achievable. We also assume that, for the equivocation rate for any two adjacent receivers $k, k+1$ for $k = 1, \ldots, K-1$ to be achievable, the $(k+1)$th receiver must have knowledge of $\mathbf{u}_{k+2}, \ldots, \mathbf{u}_K$. Then, we have

$$
\begin{aligned}
&nR_{e(k,k+1)} = H(W_k, W_{k+1}|\mathbf{Z}) \\
&\geq H(W_k, W_{k+1}|\mathbf{Z}, \mathbf{U}_{k+2}, \ldots, \mathbf{U}_K) \qquad \text{since conditioning reduces entropy} \\
&= H(W_k, W_{k+1}, \mathbf{Z}|\mathbf{U}_{k+2}, \ldots, \mathbf{U}_K) - H(\mathbf{Z}|\mathbf{U}_{k+2}, \ldots, \mathbf{U}_K) \quad \text{by (32)} \\
&\overset{(a)}{=} H(W_k, W_{k+1}, \mathbf{U}_k, \mathbf{U}_{k+1}, \mathbf{Z}|\mathbf{U}_{k+2}, \ldots, \mathbf{U}_K) - H(\mathbf{U}_k, \mathbf{U}_{k+1}|W_k, W_{k+1}, \mathbf{U}_{k+2}, \ldots, \mathbf{U}_K, \mathbf{Z}) \\
&\quad - H(\mathbf{Z}|\mathbf{U}_{k+2}, \ldots, \mathbf{U}_K) \\
&\overset{(b)}{=} H(W_k, W_{k+1}, \mathbf{U}_k, \mathbf{U}_{k+1}|\mathbf{U}_{k+2}, \ldots, \mathbf{U}_K) + H(\mathbf{Z}|W_k, W_{k+1}, \mathbf{U}_k, \mathbf{U}_{k+1}, \mathbf{U}_{k+2}, \ldots, \mathbf{U}_K) \\
&\quad - H(\mathbf{U}_k, \mathbf{U}_{k+1}|W_k, W_{k+1}, \mathbf{U}_{k+2}, \ldots, \mathbf{U}_K, \mathbf{Z}) - H(\mathbf{Z}|\mathbf{U}_{k+2}, \ldots, \mathbf{U}_K) \\
&\overset{(c)}{\geq} H(\mathbf{U}_k, \mathbf{U}_{k+1}|\mathbf{U}_{k+2}, \ldots, \mathbf{U}_K) + H(\mathbf{Z}|\mathbf{U}_k, \mathbf{U}_{k+1}, \mathbf{U}_{k+2}, \ldots, \mathbf{U}_K) - H(\mathbf{Z}|\mathbf{U}_{k+2}, \ldots, \mathbf{U}_K) \\
&\quad - H(\mathbf{U}_k, \mathbf{U}_{k+1}|W_k, W_{k+1}, \mathbf{U}_{k+2}, \ldots, \mathbf{U}_K, \mathbf{Z}) \\
&\overset{(d)}{=} H(\mathbf{U}_{k+1}|\mathbf{U}_{k+2}, \ldots, \mathbf{U}_K) + H(\mathbf{U}_k|\mathbf{U}_{k+1}, \mathbf{U}_{k+2}, \ldots, \mathbf{U}_K) \\
&\quad + [H(\mathbf{Z}|\mathbf{U}_k, \mathbf{U}_{k+1}, \mathbf{U}_{k+2}, \ldots, \mathbf{U}_K) - H(\mathbf{Z}|\mathbf{U}_{k+2}, \ldots, \mathbf{U}_K)] - H(\mathbf{U}_k, \mathbf{U}_{k+1}|W_k, W_{k+1}, \mathbf{U}_{k+2}, \ldots, \mathbf{U}_K, \mathbf{Z}) \\
&= H(\mathbf{U}_{k+1}|\mathbf{U}_{k+2}, \ldots, \mathbf{U}_K) + H(\mathbf{U}_k|\mathbf{U}_{k+1}, \mathbf{U}_{k+2}, \ldots, \mathbf{U}_K) - I(\mathbf{U}_k, \mathbf{U}_{k+1}; \mathbf{Z}|\mathbf{U}_{k+2}, \ldots, \mathbf{U}_K) \\
&\quad - H(\mathbf{U}_k, \mathbf{U}_{k+1}|W_k, W_{k+1}, \mathbf{U}_{k+2}, \ldots, \mathbf{U}_K, \mathbf{Z}),
\end{aligned} \tag{41}
$$

where (a), (b) and (d) have the first two terms by (32), and (c) has the first term by (32) and the second term by $(W_k, W_{k+1}) \to (\mathbf{U}_k \mathbf{U}_{k+1} \mathbf{U}_{k+2} \cdots \mathbf{U}_K) \to \mathbf{Z}$.

We now bound each of the terms in the last line of (41). For the first term, given $\mathbf{U}_{k+2} = \mathbf{u}_{k+2}, \ldots, \mathbf{U}_K = \mathbf{u}_K$, $\mathbf{u}_{k+1}$ has $2^{n(R_{k+1}+R'_{k+1})}$ possible values with equal probability. For the second term, given $\mathbf{U}_{k+1} = \mathbf{u}_{k+1}, \ldots, \mathbf{U}_K = \mathbf{u}_K$, $\mathbf{u}_k$ has $2^{n(R_k+R'_k)}$ possible values with equal probability. Therefore, we have

$$
\begin{cases}
H(\mathbf{U}_{k+1}|\mathbf{U}_{k+2}, \ldots, \mathbf{U}_K) = n(R_{k+1} + R'_{k+1}), \\
H(\mathbf{U}_k|\mathbf{U}_{k+1}, \mathbf{U}_{k+2}, \ldots, \mathbf{U}_K) = n(R_k + R'_k).
\end{cases} \tag{42}
$$

For the third term, it can be shown that

$$
\begin{aligned}
I(\mathbf{U}_k, \mathbf{U}_{k+1}; \mathbf{Z}|\mathbf{U}_{k+2}, \ldots, \mathbf{U}_K) &= I(\mathbf{U}_{k+1}; \mathbf{Z}|\mathbf{U}_{k+2}, \ldots, \mathbf{U}_K) + I(\mathbf{U}_k; \mathbf{Z}|\mathbf{U}_{k+1}, \mathbf{U}_{k+2}, \ldots, \mathbf{U}_K), \\
&\leq nI(U_{k+1}; Z|U_{k+2}) + nI(U_k; Z|U_{k+1}) + 2n\delta.
\end{aligned} \tag{43}
$$

For the last term, we have

$$\begin{aligned} & H(\mathbf{U}_k, \mathbf{U}_{k+1}|W_k, W_{k+1}, \mathbf{U}_{k+2}, \ldots, \mathbf{U}_K, \mathbf{Z}) \\ & = H(\mathbf{U}_{k+1}|W_{k+1}, \mathbf{U}_{k+2}, \ldots, \mathbf{U}_K, \mathbf{Z}) + H(\mathbf{U}_k|W_k, W_{k+1}, \mathbf{U}_{k+1}, \mathbf{U}_{k+2}, \ldots, \mathbf{U}_K, \mathbf{Z}) \\ & \leq H(\mathbf{U}_{k+1}|W_{k+1}, \mathbf{U}_{k+2}, \ldots, \mathbf{U}_K, \mathbf{Z}) + H(\mathbf{U}_k|W_k, \mathbf{U}_{k+1}, \mathbf{U}_{k+2}, \ldots, \mathbf{U}_K, \mathbf{Z}), \end{aligned} \tag{44}$$

where the first equality is because of the fact that $W_k$ and $\mathbf{U}_{k+1}$ are independent, and the last line is by conditioning reducing entropy. From the last line of (44), by Fano's inequality, we have

$$\begin{cases} \frac{1}{n} H(\mathbf{U}_{k+1}|W_{k+1}, \mathbf{U}_{k+2}, \ldots, \mathbf{U}_K, \mathbf{Z}) \leq \frac{1}{n}\left(1 + \bar{\lambda}^{(k+1)} \log L'_{k+1}\right) \triangleq \epsilon'_{k+1,n}, \\ \frac{1}{n} H(\mathbf{U}_k|W_k, \mathbf{U}_{k+1}, \mathbf{U}_{k+2}, \ldots, \mathbf{U}_K, \mathbf{Z}) \leq \frac{1}{n}\left(1 + \bar{\lambda}^{(k)} \log L'_k\right) \triangleq \epsilon'_{k,n}, \end{cases} \tag{45}$$

where $\epsilon'_{k,n}, \epsilon'_{k+1,n} \to 0$ for $n$ sufficiently large. We need to show that $\bar{\lambda}^{(k+1)}$ is small for $n$ sufficiently large so that (45) holds, as we already have shown that $\bar{\lambda}^{(k)}$ is small for $n$ sufficiently large. We consider the situation where the wiretapper attempts to decode $\mathbf{U}_{k+1}$ given $W_{k+1}$ by joint typicality. Then, following the same procedure to calculate $\bar{\lambda}^{(k)}$ in (39) above, we have

$$\bar{\lambda}^{(k+1)} \leq \epsilon + 2^{nR'_{k+1}} 2^{-n(I(U_{k+1};Z|U_{k+2},\ldots,U_K)-2\epsilon)}. \tag{46}$$

Since we have selected $R'_{k+1} = I(U_{k+1}; Z|U_{k+2}) - \tau = I(U_{k+1}; Z|U_{k+2}, \ldots, U_K) - \tau$, then $\bar{\lambda}^{(k+1)} \leq \epsilon$ for $\tau > 2\epsilon$ and where $\epsilon \to 0$ for $n$ sufficiently large. As a consequence, in (45), both $\bar{\lambda}^{(k)}, \bar{\lambda}^{(k+1)}$ are small for $n$ sufficiently large and (45) holds and we have

$$\frac{1}{n} H(\mathbf{U}_{k+1}|W_{k+1}, \mathbf{U}_{k+2}, \ldots, \mathbf{U}_K, \mathbf{Z}) \leq \epsilon'_{k+1,n} + \epsilon'_{k,n}. \tag{47}$$

Then, substituting (42), (43) and (47) into the last line of (41), and given $R'_{k+1}$ and $R'_k$ in (23), we have

$$nR_{e(k,k+1)} \geq nR_k + nR_{k+1} - \epsilon_{k,k+1}, \tag{48}$$

where $\epsilon_{k,k+1} = 2\tau + 2\delta + \epsilon'_{k+1,n} + \epsilon'_{k,n}$, and so security condition (8) is shown. We note that to show equivocation rates for the pair $k = 1, 2$, this can be done by following the proof above, but replacing $\mathbf{U}_1$ by $\mathbf{X}$.

Lastly, we show the proof for the equivocation sum rate $R_{e(1,\ldots,K)}$. We have

$$\begin{aligned} nR_{e(1,\ldots,K)} & = H(W_1, \ldots, W_K|\mathbf{Z}) \\ & = H(W_1, \ldots, W_K, \mathbf{Z}) - H(\mathbf{Z}) \quad \text{by (32)} \\ & \overset{(a)}{=} H(W_1, \ldots, W_K, \mathbf{U}_2, \ldots, \mathbf{U}_K, \mathbf{X}, \mathbf{Z}) - H(\mathbf{U}_2, \ldots, \mathbf{U}_K, \mathbf{X}|W_1, \ldots, W_K, \mathbf{Z}) - H(\mathbf{Z}) \\ & \overset{(b)}{=} H(W_1, \ldots, W_K, \mathbf{U}_2, \ldots, \mathbf{U}_K, \mathbf{X}) + H(\mathbf{Z}|W_1, \ldots, W_K, \mathbf{U}_2, \ldots, \mathbf{U}_K, \mathbf{X}) \\ & \quad - H(\mathbf{U}_2, \ldots, \mathbf{U}_K, \mathbf{X}|W_1, \ldots, W_K, \mathbf{Z}) - H(\mathbf{Z}) \\ & \overset{(c)}{\geq} H(\mathbf{U}_2, \ldots, \mathbf{U}_K, \mathbf{X}) + H(\mathbf{Z}|\mathbf{U}_2, \ldots, \mathbf{U}_K, \mathbf{X}) - H(\mathbf{Z}) - H(\mathbf{U}_2, \ldots, \mathbf{U}_K, \mathbf{X}|W_1, \ldots, W_K, \mathbf{Z}) \\ & \overset{(d)}{=} H(\mathbf{U}_K) + H(\mathbf{U}_{K-1}|\mathbf{U}_K) + \cdots + H(\mathbf{X}|\mathbf{U}_2, \ldots, \mathbf{U}_K) + [H(\mathbf{Z}|\mathbf{U}_2, \ldots, \mathbf{U}_K, \mathbf{X}) - H(\mathbf{Z})] \\ & \quad - H(\mathbf{U}_2, \ldots, \mathbf{U}_K, \mathbf{X}|W_1, \ldots, W_K, \mathbf{Z}) \end{aligned}$$

$$= H(\mathbf{U}_K) + H(\mathbf{U}_{K-1}|\mathbf{U}_K) + \cdots + H(\mathbf{X}|\mathbf{U}_2, \ldots, \mathbf{U}_K) - I(\mathbf{U}_2, \ldots, \mathbf{U}_K, \mathbf{X}; \mathbf{Z})$$
$$- H(\mathbf{U}_2, \ldots, \mathbf{U}_K, \mathbf{X}|W_1, \ldots, W_K, \mathbf{Z}), \quad (49)$$

where (a), (b) and (d) have the first two terms by (32), and (c) has the first term by (32) and the second term by the fact that $(W_1 \ \ldots \ W_K) \to (\mathbf{U}_K \ \ldots \ \mathbf{U}_2\mathbf{X}) \to \mathbf{Z}$.

We now bound each of the terms in the last line of (49). For the first term, $\mathbf{u}_K$ has $2^{n(R_K+R'_K)}$ possible values with equal probability. For the second to the $(K-1)$th terms, given the preceding codewords $\mathbf{u}_{k+1}, \ldots, \mathbf{u}_K$, $\mathbf{u}_k$ has $2^{n(R_k+R'_k)}$ possible values with equal probability. For the $K$th term, given all the preceding codewords, $\mathbf{x}$ has $2^{n(R_1+R'_1)}$ possible values with equal probability. As such, we have

$$H(\mathbf{U}_K) = n(R_K + R'_K), \quad (50a)$$
$$H(\mathbf{U}_{K-1}|\mathbf{U}_K) = n(R_{K-1} + R'_{K-1}), \quad (50b)$$
$$\vdots$$
$$H(\mathbf{X}|\mathbf{U}_2, \ldots, \mathbf{U}_K) = n(R_1 + R'_1). \quad (50c)$$

For the second last term, it can be shown that

$$\begin{aligned} I(\mathbf{U}_2, \ldots, \mathbf{U}_K, \mathbf{X}; \mathbf{Z}) &= I(\mathbf{U}_K; \mathbf{Z}) + I(\mathbf{U}_{K-1}; \mathbf{Z}|\mathbf{U}_K) + \cdots + I(\mathbf{X}; \mathbf{Z}|\mathbf{U}_2, \ldots, \mathbf{U}_K) \\ &\leq nI(U_K; Z) + nI(U_{K-1}; Z|U_K) + \cdots + nI(X; Z|U_2) + Kn\delta. \end{aligned} \quad (51)$$

For the last term, we have

$$\begin{aligned} H(\mathbf{U}_2, \ldots, \mathbf{U}_K, \mathbf{X}|W_1, \ldots, W_K, \mathbf{Z}) &= H(\mathbf{U}_K|W_K, \mathbf{Z}) + H(\mathbf{U}_{K-1}|W_{K-1}, W_K, \mathbf{U}_K, \mathbf{Z}) + \cdots \\ &\quad + H(\mathbf{X}|W_1, \ldots, W_K, \mathbf{U}_2, \ldots, \mathbf{U}_K, \mathbf{Z}) \\ &\leq H(\mathbf{U}_K|W_K, \mathbf{Z}) + H(\mathbf{U}_{K-1}|W_{K-1}, \mathbf{U}_K, \mathbf{Z}) + H(\mathbf{X}|W_1, \mathbf{U}_2, \ldots, \mathbf{U}_K, \mathbf{Z}) \end{aligned} \quad (52)$$

where the first equality is because of the fact that successively, we have $\mathbf{U}_K$ and $W_1, \ldots, W_{K-1}$ are independent, $\mathbf{U}_{K-1}$ and $W_1, \ldots, W_{K-2}$ are independent, and so on, and the last line is by conditioning reducing entropy. Now by applying Fano's inequality to each term in the last line of (52), we have

$$\begin{cases} \frac{1}{n}H(\mathbf{U}_K|W_K, \mathbf{Z}) \leq \frac{1}{n}\left(1 + \bar{\lambda}^{(K)} \log L'_K\right) \triangleq \epsilon'_{K,n}, \\ \frac{1}{n}H(\mathbf{U}_{K-1}|W_{K-1}, \mathbf{U}_K, \mathbf{Z}) \leq \frac{1}{n}\left(1 + \bar{\lambda}^{(K-1)} \log L'_2\right) \triangleq \epsilon'_{K-1,n}, \\ \vdots \\ \frac{1}{n}H(\mathbf{X}|W_1, \mathbf{U}_2, \ldots, \mathbf{U}_K, \mathbf{Z}) \leq \frac{1}{n}\left(1 + \bar{\lambda}^{(1)} \log L'_1\right) \triangleq \epsilon'_{1,n}. \end{cases} \quad (53)$$

where $\epsilon'_{1,n}, \ldots, \epsilon'_{K,n} \to 0$ for $n$ sufficiently large. It can be shown that $\bar{\lambda}^{(1)}, \ldots, \bar{\lambda}^{(K)} \leq 2\epsilon$ where $\epsilon \to 0$ for $n$ sufficiently large by considering the wiretapper decoding $\mathbf{X}, \ldots, \mathbf{U}_K$ given the respective associated messages $W_1, \ldots, W_K$ and the preceding codewords. This is done using the same method of joint typicality and the choice of rates for $R'_k$, $k = 1, \dots, K$ as shown in the above for the equivocation rates for $R_{e(k)}$ and $R_{e(k,k+1)}$. Thus,

$$\frac{1}{n} H(\mathbf{U}_2, \dots, \mathbf{U}_K, \mathbf{X} | W_1, \dots, W_K, \mathbf{Z}) \leq \sum_{k=1}^{K} \epsilon'_{k,n}. \tag{54}$$

Then by substituting (50a)–(50c), (51) and (54) into the last line of (49), we have, given the definitions for the rate tuple $(R'_1, \dots, R'_K)$ in (23),

$$nR_{e(1,\dots,K)} \geq n \sum_{k=1}^{K} R_k - \epsilon_{1,\dots,K}, \tag{55}$$

where $\epsilon_{1,\dots,K} = K\tau + K\delta + \sum_{k=1}^{K} \epsilon'_{k,n}$ and the security conditions in (8) are satisfied. As a result, we have shown that the equivocation rates in (8) are achievable, and hence the secret rate tuple $(R_1, \dots, R_K)$.

## IV. Proof of Converse

Here, we show the converse proof to Theorem 1. Consider a $(2^{nR_1}, \dots, 2^{nR_K}, n)$ code with error probability $P_e^{(n)}$ with the code construction so that we have the condition $(W_1 \, \cdots \, W_K) \to \mathbf{X} \to \mathbf{Y}_1 \, \cdots \, \mathbf{Y}_K \mathbf{Z}$. Then, the probability distribution on $\mathcal{W}_1 \times \cdots \times \mathcal{W}_K \times \mathcal{X}^n \times \mathcal{Y}_1^n \times \cdots \times \mathcal{Y}_K^n \times \mathcal{Z}^n$ is given by

$$p(w_1) \, \cdots \, p(w_3) p(\mathbf{x} | w_1, \dots, w_K) \prod_{i=1}^{n} p(y_{1i}, \dots, y_{Ki}, z_i | x_i). \tag{56}$$

In the following, we give the proof for the rate at the $k$th receiver. We shall also show later that the proof for the $K$th receiver can be easily obtained from this, while the proof for the first receiver requires a few additional steps.

For $k = 2, \dots, K-1$, the rate $R_k$ satisfies

$$\begin{aligned}
nR_k &= H(W_k) \leq H(W_k | \mathbf{Z}) + n\epsilon_k \qquad \text{by secrecy condition} \\
&= H(W_k | \mathbf{Z}, W_{k+1}, \dots, W_K) + I(W_k; W_{k+1}, \dots, W_K | \mathbf{Z}) + n\epsilon_k \\
&= H(W_k | W_{k+1}, \dots, W_K) - I(W_k; \mathbf{Z} | W_{k+1}, \dots, W_K) + I(W_k; W_{k+1}, \dots, W_K | \mathbf{Z}) + n\epsilon_k \\
&= I(W_k; \mathbf{Y}_k | W_{k+1}, \dots, W_K) + H(W_k | \mathbf{Y}_k, W_{k+1}, \dots, W_K) - I(W_k; \mathbf{Z} | W_{k+1}, \dots, W_K) \qquad (57) \\
&\quad + I(W_k; W_{k+1}, \dots, W_K | \mathbf{Z}) + n\epsilon_k \\
&\overset{(a)}{\leq} I(W_k; \mathbf{Y}_k | W_{k+1}, \dots, W_K) - I(W_k; \mathbf{Z} | W_{k+1}, \dots, W_K) + H(W_k | \mathbf{Y}_k, W_{k+1}, \dots, W_K) \\
&\quad + H(W_{k+1} | \mathbf{Z}) + \, \cdots \, + H(W_K | \mathbf{Z}) + n\epsilon_k \\
&\overset{(b)}{\leq} I(W_k; \mathbf{Y}_k | W_{k+1}, \dots, W_K) - I(W_k; \mathbf{Z} | W_{k+1}, \dots, W_K) + n(\delta''_k + \delta'_{k+1} + \cdots + \delta'_K + \epsilon_k), \qquad (58)
\end{aligned}$$

where (a) is by $I(W_k; W_{k+1}, \dots, W_K | \mathbf{Z}) \leq H(W_{k+1}, \dots, W_K | \mathbf{Z}) \leq H(W_{k+1} | \mathbf{Z}) + \cdots + H(W_K | \mathbf{Z})$, and (b) is by Fano's inequality which gives

$$\begin{cases}
H(W_k | \mathbf{Y}_k, W_{k+1}, \dots, W_K) \leq nR_k P_e^{(n)} + 1 \triangleq n\delta''_k, \\
H(W_{k+1} | \mathbf{Z}) \leq nR_{k+1} P_e^{(n)} + 1 \triangleq n\delta'_{k+1}, \\
\vdots \\
H(W_K | \mathbf{Z}) \leq nR_K P_e^{(n)} + 1 \triangleq n\delta'_K,
\end{cases} \tag{59}$$

where $\delta''_k,\ \delta'_{k+1},\ \ldots\ ,\ \delta'_K \to 0$ if $P_e^{(n)} \to 0$.

Expanding the first two terms of (57) by the chain rule gives

$$I(W_k;\mathbf{Y}_k|W_{k+1},\ldots,W_K) = \sum_{i=1}^{n} I(W_k;Y_{k,i}|W_{k+1},\ldots,W_K,\mathbf{Y}_k^{i-1}), \tag{60a}$$

$$I(W_k;\mathbf{Z}|W_{k+1},\ldots,W_K) = \sum_{i=1}^{n} I(W_k;Z_i|W_{k+1},\ldots,W_K,\tilde{\mathbf{Z}}^{i+1}). \tag{60b}$$

From (60a), by using the identity $I(S_1S_2;T|V) = I(S_1;T|V) + I(S_2;T|S_1V)$, we get

$$\begin{aligned} &I(W_k;Y_{k,i}|W_{k+1},\ldots,W_K,\mathbf{Y}_k^{i-1}) \\ &\quad = I(W_k,\tilde{\mathbf{Z}}^{i+1};Y_{k,i}|W_{k+1},\ldots,W_K,\mathbf{Y}_k^{i-1}) - I(\tilde{\mathbf{Z}}^{i+1};Y_{k,i}|W_k,W_{k+1},\ldots,W_K,\mathbf{Y}_1^{i-1}) \\ &\quad = I(W_k;Y_{k,i}|W_{k+1},\ldots,W_K,\mathbf{Y}_k^{i-1},\tilde{\mathbf{Z}}^{i+1}) + I(\tilde{\mathbf{Z}}^{i+1};Y_{k,i}|W_{k+1},\ldots,W_K,\mathbf{Y}_k^{i-1}) \\ &\qquad - I(\tilde{\mathbf{Z}}^{i+1};Y_{k,i}|W_k,W_{k+1},\ldots,W_K,\mathbf{Y}_k^{i-1}). \end{aligned} \tag{61}$$

Substituting this into (60a) we have,

$$\begin{cases} I(W_k;\mathbf{Y}_k|W_{k+1},\ldots,W_K) = \displaystyle\sum_{i=1}^{n} I(W_k;Y_{k,i}|W_{k+1},\ldots,W_K,\mathbf{Y}_k^{i-1},\tilde{\mathbf{Z}}^{i+1}) + \Sigma_{k,1} - \Sigma_{k,2} \\ \Sigma_{k,1} = \displaystyle\sum_{i=1}^{n} I(\tilde{\mathbf{Z}}^{i+1};Y_{k,i}|W_{k+1},\ldots,W_K,\mathbf{Y}_1^{i-1}), \\ \Sigma_{k,2} = \displaystyle\sum_{i=1}^{n} I(\tilde{\mathbf{Z}}^{i+1};Y_{k,i}|W_k,W_{k+1},\ldots,W_K,\mathbf{Y}_k^{i-1}). \end{cases} \tag{62}$$

From (60b), again by using $I(S_1S_2;T|V) = I(S_1;T|V) + I(S_2;T|S_1V)$, and substituting this into (60b), we get

$$\begin{cases} I(W_k;\mathbf{Z}|W_{k+1},\ldots,W_K) = \displaystyle\sum_{i=1}^{n} I(W_k;Z_i|W_{k+1},\ldots,W_K,\mathbf{Y}_k^{i-1},\tilde{\mathbf{Z}}^{i+1}) + \Sigma^*_{k,1} - \Sigma^*_{k,2} \\ \Sigma^*_{k,1} = \displaystyle\sum_{i=1}^{n} I(\mathbf{Y}_k^{i-1};Z_i|W_{k+1},\ldots,W_K,\tilde{\mathbf{Z}}^{i+1}), \\ \Sigma^*_{k,2} = \displaystyle\sum_{i=1}^{n} I(\mathbf{Y}_k^{i-1};Z_i|W_k,W_{k+1},\ldots,W_K,\tilde{\mathbf{Z}}^{i+1}). \end{cases} \tag{63}$$

It is known by Lemma 7 in [6] that $\Sigma_{k,1} = \Sigma^*_{k,1}$ and $\Sigma_{k,2} = \Sigma^*_{k,2}$. Therefore,

$$\begin{aligned} nR_k \le \sum_{i=1}^{n} \Big[ &I(W_k;Y_{k,i}|W_{k+1},\ldots,W_K,\mathbf{Y}_k^{i-1},\tilde{\mathbf{Z}}^{i+1}) - I(W_k;Z_i|W_{k+1},\ldots,W_K,\mathbf{Y}_k^{i-1},\tilde{\mathbf{Z}}^{i+1}) \Big] \\ &+ n(\delta''_k + \delta'_{k+1} + \cdots + \delta'_K + \epsilon_k). \end{aligned} \tag{64}$$

The terms under the summation are

$$\begin{aligned} &I(W_k;Y_{k,i}|W_{k+1},\ldots,W_K,\mathbf{Y}_k^{i-1},\tilde{\mathbf{Z}}^{i+1}) - I(W_k;Z_i|W_{k+1},\ldots,W_K,\mathbf{Y}_k^{i-1},\tilde{\mathbf{Z}}^{i+1}) \\ &= H(W_k|W_{k+1},\ldots,W_K,\mathbf{Y}_k^{i-1},Z_i,\tilde{\mathbf{Z}}^{i+1}) - H(W_k|W_{k+1},\ldots,W_K,\mathbf{Y}_k^{i-1},Y_{k,i},\tilde{\mathbf{Z}}^{i+1}) \\ &\overset{(a)}{\le} H(W_k|W_{k+1},\ldots,W_K,\mathbf{Y}_k^{i-1},Z_i,\tilde{\mathbf{Z}}^{i+1}) - H(W_k|W_{k+1},\ldots,W_K,\mathbf{Y}_k^{i-1},Y_{k,i},Z_i,\tilde{\mathbf{Z}}^{i+1}) \\ &= I(W_k;Y_{k,i}|W_{k+1},\ldots,W_K,\mathbf{Y}_k^{i-1},Z_i,\tilde{\mathbf{Z}}^{i+1}) \end{aligned}$$

$$
\begin{aligned}
&= H(Y_{k,i}|W_{k+1},\ldots,W_K,\mathbf{Y}_k^{i-1},Z_i,\tilde{\mathbf{Z}}^{i+1}) - H(Y_{k,i}|W_k,W_{k+1},\ldots,W_K,\mathbf{Y}_k^{i-1},Z_i,\tilde{\mathbf{Z}}^{i+1}) \qquad (65)\\
&= H(Y_{k,i}|W_{k+1},\ldots,W_K,\mathbf{Y}_k^{i-1},\mathbf{Y}_{k+1}^{i-1},\ldots,\mathbf{Y}_K^{i-1},Z_i,\tilde{\mathbf{Z}}^{i+1})\\
&\quad + I(Y_{k,i};\mathbf{Y}_{k+1}^{i-1},\ldots,\mathbf{Y}_K^{i-1}|W_{k+1},\ldots,W_K,\mathbf{Y}_k^{i-1},Z_i,\tilde{\mathbf{Z}}^{i+1})\\
&\quad - H(Y_{k,i}|W_k,W_{k+1},\ldots,W_K,\mathbf{Y}_k^{i-1},Z_i,\tilde{\mathbf{Z}}^{i+1})\\
&\overset{(b)}{=} H(Y_{k,i}|W_{k+1},\ldots,W_K,\mathbf{Y}_k^{i-1},\mathbf{Y}_{k+1}^{i-1},\ldots,\mathbf{Y}_K^{i-1},Z_i,\tilde{\mathbf{Z}}^{i+1}) - H(Y_{k,i}|W_k,W_{k+1},\ldots,W_K,\mathbf{Y}_k^{i-1},Z_i,\tilde{\mathbf{Z}}^{i+1})\\
&\overset{(c)}{\leq} H(Y_{k,i}|W_{k+1},\ldots,W_K,\mathbf{Y}_{k+1}^{i-1},\ldots,\mathbf{Y}_K^{i-1},Z_i,\tilde{\mathbf{Z}}^{i+1}) - H(Y_{k,i}|W_k,W_{k+1},\ldots,W_K,\mathbf{Y}_k^{i-1},\mathbf{Y}_{k+1}^{i-1},\ldots,\mathbf{Y}_K^{i-1},Z_i,\tilde{\mathbf{Z}}^{i+1})\\
&= I(W_k,\mathbf{Y}_k^{i-1};Y_{k,i}|W_{k+1},\ldots,W_K,\mathbf{Y}_{k+1}^{i-1},\ldots,\mathbf{Y}_K^{i-1},Z_i,\tilde{\mathbf{Z}}^{i+1}),
\end{aligned}
$$

where (a) and (c) are due to conditioning reducing entropy, and (b) is due to the fact that

$$
I(Y_{k,i};\mathbf{Y}_{k+1}^{i-1},\ldots,\mathbf{Y}_K^{i-1}|W_{k+1},\ldots,W_K,\mathbf{Y}_k^{i-1},Z_i,\tilde{\mathbf{Z}}^{i+1}) = 0, \qquad (66)
$$

since $\mathbf{Y}_{k+1}^{i-1}\ \cdots\ \mathbf{Y}_K^{i-1} \to W_{k+1}\ \cdots\ W_K \to \mathbf{Y}_k^{i-1} Z_i \tilde{\mathbf{Z}}^{i+1} \to Y_{k,i}$. Now, define the random variables

$$
\begin{cases}
U_{K,i} \triangleq W_K \mathbf{Y}_K^{i-1} \tilde{\mathbf{Z}}^{i+1}, \\
U_{k,i} \triangleq W_k \mathbf{Y}_k^{i-1}, \qquad \text{for } k = 2,\ldots,K-1,
\end{cases} \qquad (67)
$$

and we have the condition

$$
U_{K,i} \to\ \cdots\ \to U_{k,i} \to X_i \to Y_{k,i}\ \cdots\ Y_{K,i} \to Z_i. \qquad (68)
$$

We note that our choice of auxiliary random variables is different from Bagherikaram *et al.*, which deals with the 2-receiver degraded BC with an external wiretapper [10], and from [12], which studies the $K$-receiver degraded BC with a common message and an external wiretapper. The choice is also different, due to the presence of the wiretapper, from that of Borade *et al.* in [13] which deals with the $K$-receiver degraded BC without secrecy conditions. Thus, we have

$$
\begin{aligned}
&I(W_k,\mathbf{Y}_k^{i-1};Y_{k,i}|W_{k+1},\ldots,W_K,\mathbf{Y}_{k+1}^{i-1},\ldots,\mathbf{Y}_K^{i-1},Z_i,\tilde{\mathbf{Z}}^{i+1})\\
&= I(U_{k,i};Y_{k,i}|U_{(k+1),i},\ldots,U_{K,i},Z_i)\\
&\overset{(a)}{=} I(U_{k,i};Y_{k,i}|U_{(k+1),i},Z_i)\\
&= I(U_{k,i};Y_{k,i},Z_i|U_{(k+1),i}) - I(U_{k,i};Z_i|U_{(k+1),i})\\
&= I(U_{k,i};Y_{k,i}|U_{(k+1),i}) + I(U_{k,i};Z_i|U_{(k+1),i}) - I(U_{k,i};Z_i|U_{(k+1),i})\\
&\overset{(b)}{=} I(U_{k,i};Y_{k,i}|U_{(k+1),i}) - I(U_{k,i};Z_i|U_{(k+1),i}),
\end{aligned} \qquad (69)
$$

where (a) is due to the condition (68), and (b) is due to $I(U_{k,i};Z_i|U_{(k+1),i}) = 0$ since we have $U_{k,i} \to U_{(k+1),i} \to Z_i$. As a result, we have

$$
nR_k \leq \sum_{i=1}^{n} \left[ I(U_{k,i};Y_{k,i}|U_{(k+1),i}) - I(U_{k,i};Z_i|U_{(k+1),i}) \right] + n(\delta_k'' + \delta_{k+1}' + \cdots + \delta_K' + \epsilon_k). \qquad (70)
$$

To show the converse for $R_1$, we follow the same steps as above, but additionally we use (65) with $k = 1$ to arrive at the equivalent chain of equalities (69) above for $k = 1$. From the last line of (65), substituting for the random variables $U_{2,i}, \ldots, U_{K,i}$, we then have

$$\begin{aligned}
& I(W_1, \mathbf{Y}_1^{i-1}; Y_{1,i}|W_2, \ldots, W_K, \mathbf{Y}_2^{i-1}, \ldots, \mathbf{Y}_K^{i-1}, Z_i, \tilde{\mathbf{Z}}^{i+1}) \\
& = I(W_1, \mathbf{Y}_1^{i-1}; Y_{1,i}|U_{2,i}, \ldots, U_{K,i}, Z_i) \\
& = I(W_1; Y_{1,i}|U_{2,i}, \ldots, U_{K,i}, Z_i) + I(\mathbf{Y}_1^{i-1}; Y_{1,i}|W_1, U_{2,i}, \ldots, U_{K,i}, Z_i) \\
& \overset{(a)}{=} I(W_1; Y_{1,i}|U_{2,i}, \ldots, U_{K,i}, Z_i) \\
& \overset{(b)}{\leq} I(X_i; Y_{1,i}|U_{2,i}, Z_i) \\
& = I(X_i; Y_{1,i}, Z_i|U_{2,i}) - I(X_i; Z_i|U_{2,i}) \\
& = I(X_i; Y_{1,i}|U_{2,i}) + I(X_i; Z_i|U_{2,i}, Y_{1,i}) - I(X_i; Z_i|U_{2,i}) \\
& \overset{(c)}{=} I(X_i; Y_{1,i}|U_{2,i}) - I(X_i; Z_i|U_{2,i})
\end{aligned} \tag{71}$$

where (a) is by the second term $I(\mathbf{Y}_1^{i-1}; Y_{1,i}|W_1, U_{2,i}, \ldots, U_{K,i}, Z_i) = 0$ since $\mathbf{Y}_1^{i-1} \rightarrow W_1 U_{2,i} \cdots U_{K,i} Z_i \rightarrow Y_{1,i}$, (b) is by $Y_{1,i} \rightarrow X_i \rightarrow W_1$ and by the Markov condition (68), and (c) is by the second term $I(X_i; Z_i|U_{2,i}, Y_{1,i}) = 0$ since $X_i \rightarrow U_{2,i} Y_{1,i} \rightarrow Z_i$. Thus, we have

$$nR_1 \leq \sum_{i=1}^{n} [I(X_i; Y_{1,i}|U_{2,i}) - I(X_i; Z_i|U_{2,i})] + n(\delta_1'' + \delta_2' + \cdots + \delta_K' + \epsilon_1). \tag{72}$$

The proof for $R_K$ is easily obtained using the above approach, only without conditioning on $W_{k+1}, \ldots, W_K$ in the second line of (57). This results in

$$nR_K \leq \sum_{i=1}^{n} [I(U_{K,i}; Y_{K,i}) - I(U_{K,i}; Z_i)] + n(\delta_K'' + \epsilon_K). \tag{73}$$

Now, we introduce the random variable $G$, which is uniformly distributed among the integers $\{1, 2, \ldots, n\}$ and is independent of all other random variables. Define the following auxiliary random variables

$$U_K = (G, U_{K,G}), \tag{74a}$$

$$U_{K-1} = (G, U_{K-1,G}), \tag{74b}$$

$$\vdots$$

$$X = X_G, \tag{74c}$$

$$Y_1 = Y_{1,G}, \tag{74d}$$

$$\vdots$$

$$Y_K = Y_{K,G}, \tag{74e}$$

$$Z = Z_G. \tag{74f}$$

Then (70), (72), (73) become

$$R_K \leq I(U_K;Y_K) - I(U_K;Z), \tag{75a}$$

$$R_k \leq I(U_k;Y_k|U_{k+1}) - I(U_k;Z|U_{k+1}), \quad \text{for } k = 2,\ldots,K-1, \tag{75b}$$

$$R_1 \leq I(X;Y_1|U_2) - I(X;Z|U_2), \tag{75c}$$

and the converse to Theorem 1 is proved.

## V. Conclusion

We have presented a new secrecy capacity region for the degraded $K$-receiver BC with private messages in the presence of a wiretapper which generalizes previous work which dealt with 2-receiver BCs. In the direct proof we have used superposition coding and Wyner's random code partitioning instead of binning to show the achievable rate tuples. We have provided error probability analysis and equivocation calculation for the general $k$th receiver. In the converse proof we have used a new definition for the auxiliary random variables which is different from either the 2-receiver BC with a wiretapper, or the more recently studied $K$-receiver BC with common message and wiretapper, or the $K$-receiver BC without wiretapper cases.

## References

[1] Y. Liang and H. Poor, "Multiple-access channels with confidential messages," *IEEE Trans. Info. Theory*, vol. 54, no. 3, pp. 976–1002, Mar. 2008.

[2] L. Lai and H. E. Gamal, "The relay-eavesdropper channel: Cooperation for secrecy," *IEEE Trans. Info. Theory*, vol. 54, no. 9, pp. 4005–4019, Sep. 2008.

[3] M. Bloch, J. Barros, M. Rodrigues, and S. McLaughlin, "Wireless information-theoretic security," *IEEE Trans. Info. Theory*, vol. 54, no. 6, pp. 2515–2534, Jun. 2008.

[4] Y. Liang, H. Poor, and S. Shamai, "Secure communication over fading channels," *IEEE Trans. Info. Theory*, vol. 54, no. 6, pp. 2470–2492, Jun. 2008.

[5] R. Liu and H. Poor, "Secrecy capacity region of a multi-antenna Gaussian broadcast channel with confidential messages," *IEEE Trans. Info. Theory*, submitted for publication, Sep. 2007.

[6] I. Csiszár and J. Körner, "Broadcast channels with confidential messages," *IEEE Trans. Info. Theory*, vol. 24, no. 3, pp. 339–348, Mar. 1978.

[7] A. D. Wyner, "The wire-tap channel," *Bell Syst. Tech. J.*, vol. 54, no. 8, pp. 1355–1387, 1975.

[8] R. Liu, I. Marić, P. Spasojević, and R. Yates, "Discrete memoryless interference and broadcast channels with confidential messages: Secrecy rate regions," *IEEE Trans. Info. Theory*, vol. 54, no. 6, pp. 2493–2507, Jun. 2008.

[9] J. Xu, Y. Cao, and B. Chen, "Capacity bounds for broadcast channels with confidential messages," *IEEE Trans. Info. Theory*, submitted for publication, May 2008.

[10] G. Bagherikaram, A. S. Motahari, and A. K. Khandani, "Secrecy rate region of the broadcast channel," *IEEE Trans. Info. Theory*, submitted for publication, Jul. 2008.

[11] L. C. Choo and K. K. Wong, "Three-receiver broadcast channel with confidential messages," *IEEE Info. Theory Workshop, ITW 2009, Volos, Greece,* submitted for publication, Dec. 2008.

[12] E. Ekrem and S. Ulukus, "Secrecy capacity of a class of broadcast channels with an eavesdropper," *EURASIP J. Wireless Commun. and Net.*, submitted for publication, Dec. 2008.

[13] S. Borade, L. Zheng, and M. Trott, "Multilevel broadcast networks," in Proc. *Int. Sym. Info. Theory*, Nice, France, 24–29, 2007, pp. 1151–1155.

[14] P. Bergmans, “Random coding theorem for broadcast channels with degraded components,” *IEEE Trans. Info. Theory*, vol. 19, no. 2, pp. 197–207, Mar. 1973.

[15] T. Cover and J. Thomas, *Elements of Information Theory*, 2nd ed., Wiley, 2006.